\documentclass[]{elsarticle}

\usepackage{lineno,hyperref}
\modulolinenumbers[5]

\journal{Physics Letters B}

\usepackage[super,compress]{cite}
\usepackage{graphicx}

\usepackage{braket}
\usepackage{multirow}
\usepackage{dsfont}
\usepackage{mathtools}
\usepackage{xcolor}
\usepackage[normalem]{ulem}

\newcommand{\bA}{\boldsymbol{A}}
\newcommand{\bt}{\boldsymbol{t}}

\renewcommand{\vec}[1]{\mbox{\boldmath$#1$\unboldmath}}

\begin{document}

\begin{frontmatter}

  \title{The chiral magnetic effect and the chiral spin symmetry in QCD above $T_c$}
  \author{L. Ya. Glozman}
  \address{Institute of Physics, University of Graz, A-8010 Graz, Austria}

\begin{abstract}
  The chiral magnetic effect (CME) is an exact statement that
  connects via the axial anomaly the electric current in a 
  system consisting of interacting fermions and gauge field
  with chirality imbalance that is put into a strong external magnetic
  field. Experimental search of the magnetically induced current
   in QCD in heavy ion collisions above a pseudocritical temperature hints,
   though not yet conclusive, that the
  induced current is either small or vanishing. This would imply that
  the chirality imbalance in QCD above $T_c$ that could be generated via topological
  fluctuations is at most very small. Here we present the most general
  reason for absence (smallness) of the chirality imbalance in QCD
  above $T_c$. It was recently found on the lattice that QCD above $T_c$ is approximately
  chiral spin  (CS) symmetric with the symmetry breaking at the level
  of a few percent. The CS transformations mix the right-
  and left-handed components of quarks. Then an exact CS
  symmetry would require absence of any chirality imbalance. Consequently
  an approximate CS symmetry admits at most a very small
  chirality imbalance in QCD above $T_c$. Hence the absence or smallness of an
  magnetically induced current observed in heavy ion collisions could be considered
  as  experimental evidence for emergence of the CS symmetry above $T_c$.  
\end{abstract}


\end{frontmatter}


\section{Introduction}

According to the Atiah-Singer theorem local topological fluctuations
of the gluonic field in Euclidean space with nonzero topological charge induce creation 
of chiral quarks  such that the number of the right-handed quarks
exceeds the number of the left-handed quarks by the topological charge.
This process produces a  chirality imbalance. If such a system
is put into a strong external magnetic field, then this process should give
rise to the magnetically induced electric current. This phenomenon is
called the chiral magnetic effect (CME) \cite{K1,K2} and the exact relation
between the external magnetic field and magnetically induced current based on axial anomaly
is

 \begin{equation}
\label{cme}
 \vec{J} =  \sigma_5 \vec{B}, 
 \end{equation}

\noindent
where the chiral magnetic conductivety $\sigma_5$ is expressed in terms
of the chiral chemical potential 
 as
 
 \begin{equation}
\label{pot}
  \sigma_5 = N_c \sum _f \frac {Q^2_f e^2}{2\pi^2} \mu_5
 \end{equation}

\noindent
with $Q_f e$ being the electric charge of the quark with
the flavor $f$ and

\begin{equation}
\label{mu}
\mu_5 = \frac{\mu_R - \mu_L}{2}
\end{equation}

\noindent
being the axial chemical potential, i.e a difference of the chemical potentials
for the right- and  left-handed quarks. The axial chemical potential parametrises
the  chirality imbalance induced by local topological fluctuations.

Experimental search of CME at RHIC and LHC suggests that
 the magnetically induced current is at most very small, though
 there are many uncertanties and direct extraction of the current
 from data is impossible \cite{W1,W2,W3}. If the magnetic field that is
 formed during collision of two nuclei is sufficiently large and
 at the same time a magnetically induced electric current vanishes or small,
 as hinted by experimental data, then one infers from (\ref{cme})
 that chirality imbalance in QCD is either absent or very small.
 This obviously requires a convincing explanation.

 It was recently suggested \cite{G1} and then observed on the lattice  \cite{R1,R2,R3}
 that QCD at RHIC and LHC temperatures is approximately chiral spin (CS)
 symmetric \cite{G2,G3}. This symmetry, that includes chiral symmetry $U(1)_A$ as a subgroup, is
 not a symmetry of the Dirac Lagrangian but is a symmetry of the Lorentz-invariant fermion charge and consequently is a
 symmetry of the chromoelectric interactions in QCD in a given reference frame.
 The chromomagnetic interaction as well as the quark kinetic term break this symmetry.
 Observation of this symmetry in $T_c - 3 T_c$ interval implies that a dominant
 physics at this temperatures is due to the chromoelectric interaction between
 chirally symmetric quarks, that are bound into the color-singlet objects
 ("strings"), and a contribution of the chromomagnetic interaction as well
 as of the quark kinetic energy is at least much smaller. The CS symmetry is not
 a symmetry of the Dirac Lagrangian and consequently implies that there are
 no free deconfined quarks.
 
 A salient feature of the CS-transformations is that they mix the right-
 and left-handed quarks. Then  exact CS symmetry in QCD above $T_c$
 would require that the chirality imbalance should be absent since this
 imbalance is proportional the difference of number of the right- and left-handed quarks.
 Hence an exact CS-symmetry would require vanishing of the magnetically
 induced electric current in heavy ion collisions. In reality this
 symmetry is not exact and is broken at a few percent level. Then this
 approximate CS symmetry admits only a very small magnetically
 induced current. 
 Stated otherwise a nonobservation or observation of only small magnetically
 induced current provided that the external  magnetic field is sufficiently
 large could be considered as an experimental evidence of the
 CS symmetry observed on the lattice.
 
 This short paper is structured in the following way.
 In  sections 2 and 3 we overview already known results
 about CS symmetry and its observation on the lattice at
 high temperatures. This will allow   to avoid
 reading the preceeding papers. Then in section 4 we
 present the key argument of this paper and will conclude
 in section 5.

\section{Chiral spin symmetry} 
 
 The chiral spin $SU(2)_{CS}$  transformation 
 was defined in ref. \cite{G2} as a transformation that rotates in the space of the
 right- and left-handed Weyl spinors

 \begin{equation}
\label{W}
 \left(\begin{array}{c}
R\\
L
\end{array}\right) \rightarrow   \exp \left(i  \frac{\varepsilon^n \sigma^n}{2}\right) \left(\begin{array}{c}
R\\
L
\end{array}\right)\; .
\end{equation}
   In terms of the Dirac spinors $\psi$ this transformation can be written via $\gamma$-matrices \cite{G3}

\begin{equation}
\label{V-defsp}
  \psi \rightarrow  \psi^\prime = \exp \left(i  \frac{\varepsilon^n \Sigma^n}{2}\right) \psi = \exp \left(i  \frac{\varepsilon^n \sigma^n}{2}\right) \left(\begin{array}{c}
R\\
L
\end{array}\right)\; ,
\end{equation}

\noindent
where the generators $\Sigma^n$ of the four-dimensional reducible
representation are

\begin{equation}
 \Sigma^n = \{\gamma_0,-i \gamma_5\gamma_0,\gamma_5\}.
\label{SIGCS}
\end{equation}
The $su(2)$ algebra is automatically satisfied for these three
generators,

\begin{equation}
[\Sigma^a,\Sigma^b]=2i\epsilon^{abc}\Sigma^c.
\label{algebra}
\end{equation}
The $U(1)_A$ group is a subgroup of $SU(2)_{CS}$.

In Euclidean space with the $O(4)$ symmetry all four directions are 
equivalent and one can use any Euclidean  hermitian $\gamma$-matrix $\gamma_k$, $k=1,2,3,4$ instead
of Minkowskian $\gamma_0$:

\begin{equation}
 \Sigma^n = \{\gamma_k,-i \gamma_5\gamma_k,\gamma_5\},
\label{SIGCS}
\end{equation} 

\begin{equation}
\gamma_i\gamma_j + \gamma_j \gamma_i =
2\delta^{ij}; \qquad \gamma_5 = \gamma_1\gamma_2\gamma_3\gamma_4.
\label{gamma}
\end{equation}
The $su(2)$ algebra 
is satisfied with any $k=1,2,3,4$.

The direct product of the $SU(2)_{CS}$ group with the flavor group
$SU(2)_{CS} \times SU(N_F)$ can be extended to a $SU(2N_F)$ group.
 This group includes the chiral
symmetry  $SU(N_F)_L \times SU(N_F)_R \times U(1)_A$ as a subgroup.
The  $SU(2N_F)$ transformations  are given by

\begin{equation}
\psi \rightarrow  \psi^\prime = \exp\left(i \frac{\epsilon^m T^m}{2}\right) \psi, 
\end{equation}

\noindent
with $m=1,2,...,(2N_F)^2-1$ and the set of $(2N_F)^2-1$ generators being

\begin{align}
T^m=\{
(\tau^a \otimes {1}_D),
({1}_F \otimes \Sigma^n),
(\tau^a \otimes \Sigma^n)
\}
\end{align}
where  $\tau$  are the flavor generators (with the flavor index $a$) and $n=1,2,3$ is the $SU(2)_{CS}$ index.

The fundamental
vector of $SU(2N_F)$ at $N_F=2$ is

\begin{equation}
\Psi =\begin{pmatrix} u_{\textsc{R}} \\ u_{\textsc{L}}  \\ d_{\textsc{R}}  \\ d_{\textsc{L}} \end{pmatrix}. 
\end{equation}
\noindent

The $SU(2)_{CS}$ and $SU(2N_F)$ groups are not
symmetries of the Dirac Lagrangian. At the same time they are symmetries
of the Lorentz-invariant fermion charge
\begin{equation}
Q = \int d^3x \bar \psi(x) \gamma_0 \psi(x) = \int d^3x  
\psi^\dagger(x)  \psi(x).
\label{Q}
\end{equation}

\noindent
This salient  feature allows us to use the 
$SU(2)_{CS}$ and $SU(2N_F)$ symmetries to distinguish the
chromoelectric and chromomagnetic interactions in a given reference frame
because  the chromoelectric interaction is influenced only by
the color charge while the chromomagnetic interaction is dictated by
the spatial current. The latter current is not 
$SU(2)_{CS}$ and $SU(2N_F)$ symmetric. 

More specifically  the (chromo)electric and (chromo)magnetic fields
in Minkowski space in a given reference frame are
 different fields.
Interaction of  fermions with the gauge field in Minkowski space-time
can be split in a given reference frame into temporal and spatial parts:

\begin{equation}
\overline{\psi}   \gamma^{\mu} D_{\mu} \psi = \overline{\psi}   \gamma^0 D_0  \psi 
  + \overline{\psi}   \gamma^i D_i  \psi .
\label{cl}
\end{equation}
\noindent
The covariant derivative $D_{\mu}$  includes
interaction of the matter field $\psi$ with the  gauge field $\bA_\mu$,

\begin{equation}
D_{\mu}\psi =( \partial_\mu - ig \frac{\bt \cdot \bA_\mu}{2})\psi.
\end{equation}
The temporal term contains  interaction of the color-octet
 charge density 

\begin{equation}
\bar \psi (x)  \gamma^0  \frac{\bt}{2} \psi(x) = \psi (x)^\dagger  \frac{\bt}{2} \psi(x)
\label{den}
\end{equation}
with the chromoelectric  
part of the gluonic field. 
It is invariant  under 
 $SU(2)_{CS}$  and  $SU(2N_F)$ since it is invariant under any unitary transformation
 that acts in Dirac and/or flavor spaces. The $SU(2)_{CS}$ transformations
defined  via the Euclidean
Dirac matrices can be identically applied to Minkowski Dirac spinors without
any modification of the generators.
 
 The spatial part contains the quark kinetic term
and  the interaction of the spatial current $\bar \psi (x)  \gamma^i  \frac{\bt}{2} \psi(x)$   
with the chromomagnetic field.  It breaks 
 $SU(2)_{CS}$ and $SU(2N_F)$.   We conclude that  interaction
 of the chromoelectric and  chromomagnetic components
 of the gauge field with quarks in a given reference frame can be distinguished
 by symmetry. 
 
Of course, in order to discuss the chromoelectric and chromomagnetic
components of the gluonic field
one needs to fix a reference frame. The hadron invariant mass 
is  the rest frame energy. Consequently, to address physics
of hadron mass  one should discuss energy in the hadron rest frame.
At high temperatures the Lorentz invariance is broken and 
a natural frame  is the medium rest frame.

\section{Emergence of chiral spin and $SU(2N_F)$
symmetries in QCD above $T_c$}

\begin{figure}
\centering
\includegraphics[angle=0,width=0.45\linewidth]{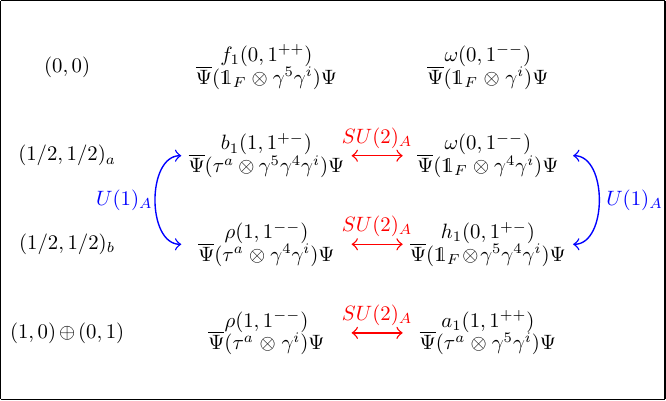}
\includegraphics[angle=0,width=0.45\linewidth]{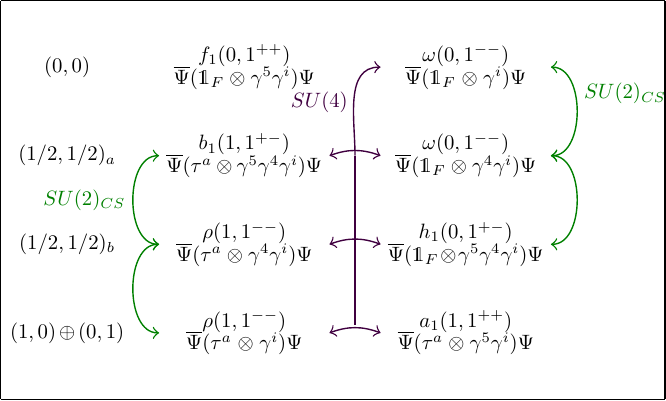}
\caption{Transformations between $J=1$ operators, $i=1,2,3$.
The left columns indicate the $SU(2)_L \times SU(2)_R$ 
representation for every
operator. Red and blue arrows connect operators which transform into 
each other under $SU(2)_L \times SU(2)_R$ and $U(1)_A$, respectively.
Green arrows connect operators that belong to
$SU(2)_{CS}$, $k=4$  triplets. Purple arrow shows the $SU(4)$
15-plet. The $f_1$ operator is is a singlet of $SU(4)$.
The Fig. is from Ref. \cite{G3}.}
\label{F1}
\end{figure}

Above the chiral restoration pseudocritical temperature $T_c$
one apriori expects  in observables chiral 
$SU(2)_L \times SU(2)_R$ symmetry  because the quark condensate vanishes
during a very smooth crossover at temperatures between 100 and 200 MeV. A pseudocritical temperature $T_c$ determined
from the chiral susceptibility is
 around 155 MeV in $N_F=2+1$ QCD \cite{F}. This symmetry above
the crossover is evidenced by degeneracy of correlators connected
by the chiral transformation. While the axial anomaly is a pertinent
property of QCD its effect is determined by the topological charge
density. There are strong indications from the lattice that the
$U(1)_A$ symmetry is also effectively restored above $T_c$ \cite{JLQCD1,JLQCD3}
which suggests that the local topological fluctuations in Euclidean space
in QCD are at least very strongly suppressed above $T_c$. It is a matter of the present
debates whether the $U(1)_A$ restoration happens at the same temperature as of
$SU(2)_L \times SU(2)_R$ or at a slightly higher temperature \cite{hotQCD}. The
$U(1)_A$ restoration is evidenced by  degeneracy of correlators connected by the
$U(1)_A$ transformation. $SU(2)_L \times SU(2)_R$ and $U(1)_A$ transformation
properties of the $J=1$ operators age given in the left panel of Fig. \ref{F1}.

In the right panel of the same Fig. we present transformation
properties of the same operators with respect to $SU(2)_{CS}$
and $SU(4)$ \cite{G3}. If one observes on the lattice  degeneracy
of correlators that are connected by $SU(2)_{CS}$ and $SU(4)$ that
would be a signal for emergence of these symmetries.

\begin{figure}
  \centering
  \includegraphics[scale=0.45]{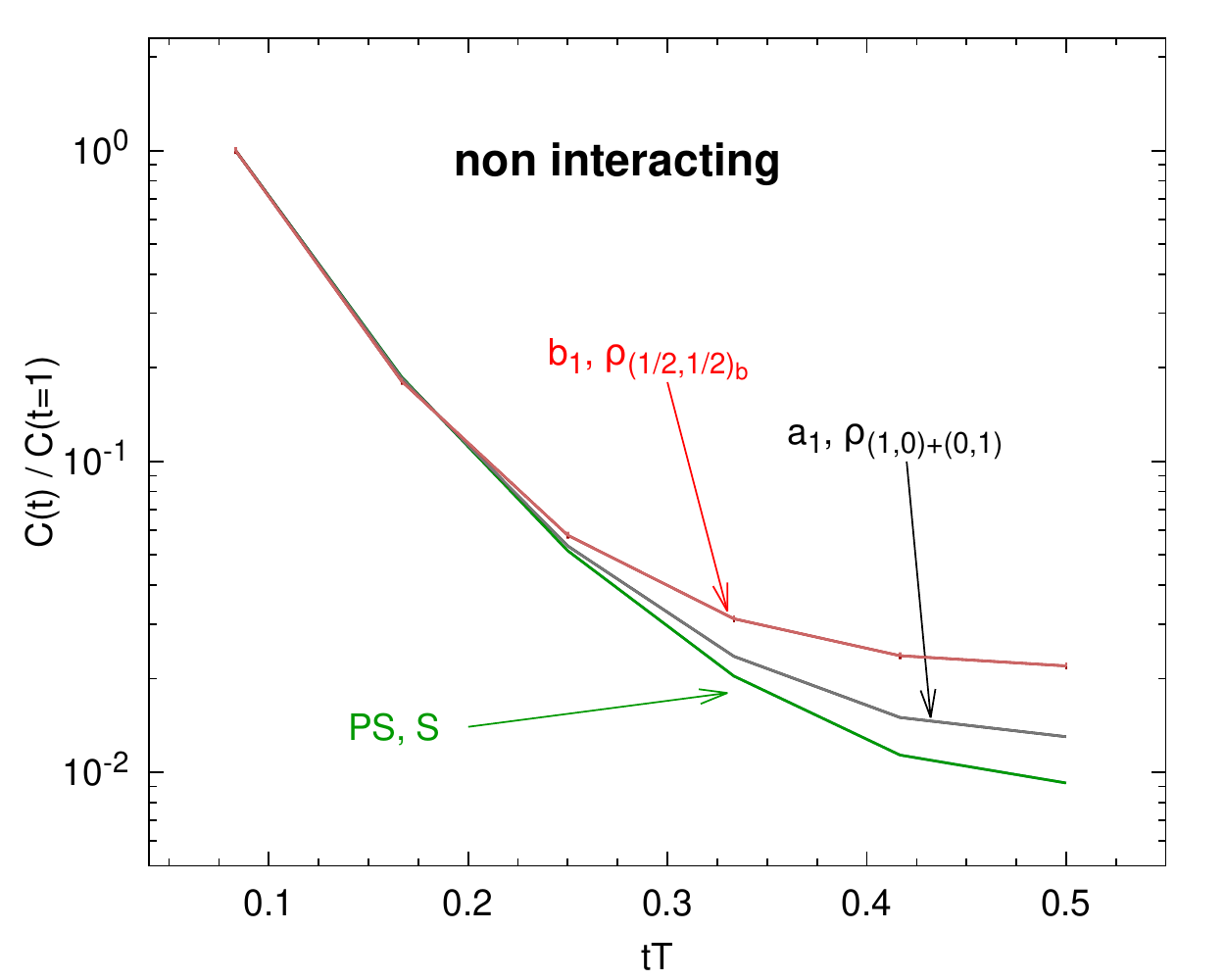} 
  \includegraphics[scale=0.45]{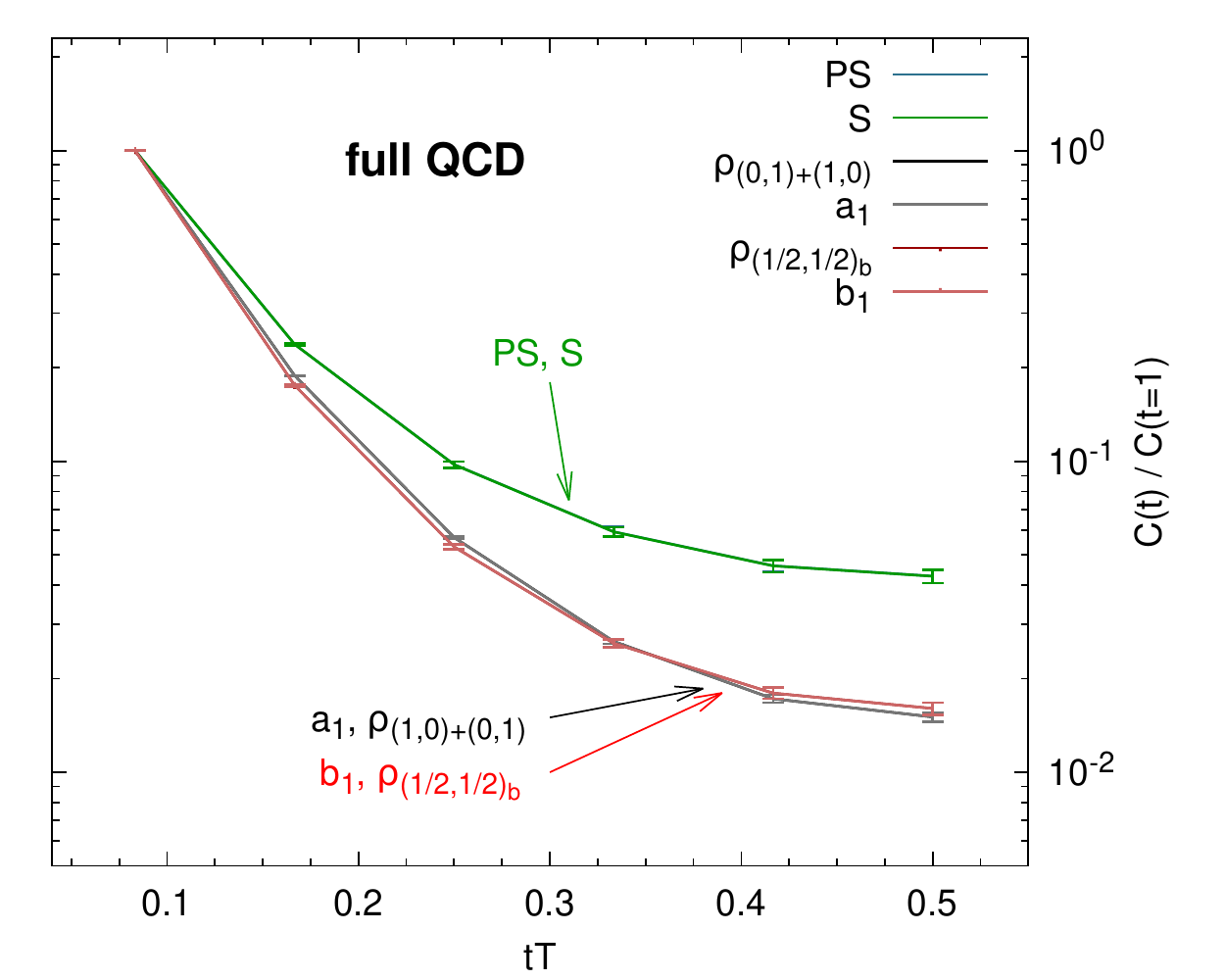} 
\caption{ Temporal correlation functions for $12 \times 48^3$
lattices. The l.h.s. shows correlators calculated with free
noninteracting quarks with manifest $U(1)_A$  and $SU(2)_L \times SU(2)_R$
symmetries. The r.h.s. presents full QCD results at a temperature $1.2 T_c$,
which shows multiplets of all  $U(1)_A$, $SU(2)_L \times SU(2)_R$, $SU(2)_{CS}$  and $SU(4)$ groups. The Fig. is from Ref. \cite{R3}.
}
\label{tcorr}
\end{figure}

On the r.h.s.  of Fig.~\ref{tcorr} we show temporal correlators

\begin{equation}
C_\Gamma(t) = \sum\limits_{x, y, z}
\braket{\mathcal{O}_\Gamma(x,y,z,t)
\mathcal{O}_\Gamma(\mathbf{0},0)^\dagger},
\label{eq:momentumprojection}
\end{equation}
at a temperature $T = 1.2 T_c$  calculated in $N_F=2$
QCD with a chirally symmetric Dirac operator \cite{R3}.
Here $\mathcal{O}_\Gamma(x,y,z,t)$ is an operator that creates a
quark-antiquark pair  with fixed quantum numbers. Summation over
$x,y,z$ projects out the  rest frame. 

Correlators of the isovector scalar (S) and isovector pseudoscalar (PS)
operators  are connected by the $U(1)_A$ transformation and
their degeneracy indicates restoration of this symmetry. If there
is a tiny splitting of the S and PS correlators then it should be so small
so that it cannot be seen in the present lattice data. This strongly suggests
that the topological transitions are at least severely suppressed above
$T_c$. An approximate degeneracy of the $a_1$, $b_1$, $\rho_{(1,0)+(0,1)}$
and  $\rho_{(1/2,1/2)_b}$ correlators indicates emergent $SU(2)_{CS}$  and $SU(4)$
symmetries. Their breaking is estimated at the level of less than $ 5\%$.

A similar multiplet structure is seen in spatial correlators in the temperature
range $T_c - 3 T_c$ \cite{R1,R2}.

On the l.h.s of Fig.~\ref{tcorr} we present correlators
calculated with noninteracting quarks on the same lattice.
They represent a QGP at a very high temperature where due to asymptotic freedom
the quark-gluon interaction can be neglected.
Dynamics of free quarks are governed by the Dirac
equation and only $U(1)_A$ and $SU(2)_L \times SU(2)_R$ chiral symmetries
exist. A qualitative difference between the pattern on the l.h.s.   and the pattern on the r.h.s of Fig.~\ref{tcorr}
is remarkable.  The temporal correlators
are directly connected to  measurable spectral density.
$SU(2)_{CS}$ and $SU(4)$ symmetries of the t-correlators imply the same 
symmetries of spectral densities.

\section{Chiral spin transformation properties of the axial chemical potential}

Now we arrive at the key point of this note: What are implications
of the emerging chiral spin symmetry
above $T_c$ 
on axial chemical potential term in effective action?

The quark chemical potential  $ \mu \psi (x)^\dagger \psi(x)$  and the axial chemical
potential $ \mu_5 \psi (x)^\dagger \gamma_5 \psi(x)$
terms can be present in the  QCD action,

\begin{equation}
S = \int_{0}^{\beta} dt \int d^3x
\overline{\psi}  [ \gamma_{\mu} D_{\mu} + \mu \gamma_4 +  \mu_5 \gamma_4 \gamma_5 +   m] \psi,
\end{equation}
\noindent
 in the  $SU(2)_{CS}$  and $SU(2N_F)$ symmetric regime only if they are invariant
 with respect to the chiral spin transformations, i.e. are the chiral spin singlets. The 
 chemical potential term is indeed invariant, i.e. transforms into 
 itself upon the $SU(2)_{CS}$ transformation (\ref{V-defsp})

 \begin{equation}
 \psi (x)^\dagger \psi(x) \longrightarrow \psi (x)^\dagger \psi(x).
\end{equation}
\noindent
 The axial chemical potential
 $ \psi (x)^\dagger \gamma_5 \psi(x)$ term is not invariant under (\ref{V-defsp})
 and transforms into a superposition of three terms:

 \begin{equation}
 \psi (x)^\dagger \gamma_5 \psi(x) \longrightarrow  \alpha \psi (x)^\dagger \gamma_4 \psi(x)
 + \beta \psi (x)^\dagger \gamma_4 \gamma_5 \psi(x) + \gamma \psi (x)^\dagger \gamma_5 \psi(x),
\end{equation}
\noindent
 i.e. it transforms under the triplet representation of $SU(2)_{CS}$ with
 coefficients $\alpha,\beta,\gamma$ being determined by three rotation angles.
 Then such a term {\it is not allowed} in a $SU(2)_{CS}$-symmetric theory.
 In other words, the $SU(2)_{CS}$ symmetry prohibiths  existence of a finite
 axial chemical potential.
 
 The same statement can be understood with a less formal language. The axial chemical
 potential parametrizes an excess of the right-handed quarks over the left-handed
 quarks. The $SU(2)_{CS}$ transformation mixes the right-
 and left-handed quarks, see (\ref{W}). Consequently a given  finite excess of the right-
 over the left-handed quarks cannot be $SU(2)_{CS}$-symmetric. Only a vanishing axial
 chemical potential is consistent with chiral spin symmetry.
 
 We conclude that if the emerged chiral spin symmetry were exact, then
 the chiral magnetic conductivety $\sigma_5$ in (\ref{cme}) must vanish.
 {\it Exact $SU(2)_{CS}$ symmetry requires vanishing of the magnetically
 induced electrical current even if the external magnetic field is large.}
 
 Of course, in reality the chiral spin symmetry above $T_c$ is not exact: 
 it is broken at a few percent level. Then the magnetically induced
 electric current can be only very small as it would originate only from the
 CS symmetry breaking contributions. The topological fluctuations with
 a nonzero topological charge do indeed break the CS symmetry. However their role
 in the QCD dynamics above $T_c$ can be only very small because of a very good
 CS symmetry.

This result explains a vanishing or very small magnetically induced electric
current above $T_c$ as it is suggested by the present experimental data.
Stated otherwise, if future experiments confirm a smallness or absence
of the magnetically induced current provided that the transient external
magnetic field is sufficiently strong, it would be an experimental evidence
of the chiral spin symmetry above $T_c$.

 \section{Conclusions}

  A   formation of multiplets in  correlators described by the chiral spin 
$SU(2)_{CS}$ and $SU(4)$ groups \cite{G2,G3} in the range $T_c - 3 T_c$ was observed on the lattice \cite{R1,R2,R3}.   
These symmetries include the chiral $U(1)_A$ and $SU(2)_L \times SU(2)_R$  as subgroups.  These are not symmetries of the free Dirac action
and they are not consistent with free deconfined quarks. In  the medium rest frame
the  chromoelectric interaction is invariant under both $SU(2)_{CS}$ and $SU(4)$ transformations,
while the chromomagnetic interaction as well as the quark kinetic term break them. 

The emergence of these symmetries in the $T_c$ - $3 T_c$  window
suggests that the chromomagnetic field disappears or is strongly
suppressed, while the confining chromoelectric
field is still active.
This implies that the
physical degrees of freedom are chirally symmetric quarks bound by the
chromoelectric interaction into color-singlet objects without chromomagnetic effects.
This regime of QCD was named as a "stringy fluid".

These symmetries are broken at a few percent level.

If the $SU(2)_{CS}$ and $SU(4)$ symmetries were exact it would require a vanishing
of the chirality imbalance. Consequently the magnetically induced current would
exactly vanish even if the external magnetic field be very strong. Then a tiny magnetically
induced current can be only related to the CS symmetry breaking dynamics which is
however much less important than a confining chromoelectric interaction that binds the
chirally symmetric quarks into color singlet objects. This conclusion is drawn from the
smallness of the CS symmetry breaking.

Confirmation in experiments of  smallness of the magnetically induced
current or its absence provided that the external magnetic field is sufficiently strong could
be considered as an experimental verification of the chiral spin and $SU(2N_F)$ symmetries
in QCD above $T_c$.

The author is thankful to S. Voloshin for discussion of the present experimental
situation as well as to M. Chernodub, T. Cohen and C. Lang for their reading of the ms
and discussions.

\end{document}